\documentclass[useAMS]{mn2e}
\usepackage{psfig}

\usepackage{times}

\title[The blazar sequence: a new perspective]
{The blazar sequence: a new perspective}

\author[G. Ghisellini and F. Tavecchio]
{G. Ghisellini$^{1}$\thanks{E--mail: gabriele.ghisellini@brera.inaf.it}  
and F. Tavecchio$^{1}$ \\
$^{1}$INAF--Osservatorio Astronomico di Brera, via E. Bianchi 46, I-23807 Merate, Italy \\
}
\begin{document}

% \date{Accepted 1988 December 15. Received 1988 December 14; 
% in original form 1988 October 11}

%\pagerange{\pageref{firstpage}--\pageref{lastpage}} \pubyear{2007}

\maketitle

\label{firstpage}

\begin{abstract}

We revisit the so called ``blazar sequence'', which connects the
observed bolometric luminosity to the shape of the spectral energy
distribution (SED) of blazars.  We propose that the power of the jet
and the SED of its emission are linked to the two main parameters of
the accretion process, namely the mass of the black hole and the
accretion rate. We assume: i) that the jet kinetic power is
proportional to the mass accretion rate; ii) that most of the jet
dissipation takes place at a distance proportional to the black hole
mass; iii) that the broad line region exists only above a critical
value of the disk luminosity, in Eddington units, and iv) that the
radius of the broad line region scales as the square root of the
ionising disk luminosity.  These assumptions, motivated by existing
observations or by reasonable theoretical considerations, are
sufficient to uniquely determine the SED of all blazars.
This framework accounts for the existence of ``blue quasars", i.e.
objects with broad emission lines but with SEDs resembling those of
low luminosity high energy peaked BL Lac objects, as well as the
existence of relatively low luminosity ``red" quasars.  Implications
on the possible evolution of blazars are briefly discussed.  This
scenario can be tested quite easily once the {\it AGILE} and
especially the {\it GLAST} satellite observations, coupled with
information in the optical/X--ray band from {\it Swift}, will
allow the knowledge of the entire SED of hundreds (and possibly
thousands) blazars.
\end{abstract}

\section{Introduction}

Fossati et al. (1998) studied three complete sample of blazars:
the Einstein Slew survey (Elvis et al. 1992), the 1--Jy samples of BL Lacs
(K\"uhr et al. 1981), and the flat--spectrum radio--loud quasars (FSRQs)
extracted by Padovani \& Urry (1992) from the 2--Jy sample 
of Wall \& Peacock (1985).
The total number of studied blazars was 126, and 33 of these were detected 
in $\gamma$--rays by the EGRET instrument onboard the 
{\it Compton Gamma Ray Observatory}.
These blazars were divided into radio luminosity bins, and the luminosity
in selected bands was averaged to form the SED representative of the blazars 
in each bin.
It turned out that the division into radio luminosity bins well matched the
division into bins of bolometric luminosity, and that all spectra could 
be described  by two broad peaks, in a $\nu F_\nu$ representation, the first 
at mm/soft X--rays frequencies, the second in the MeV--GeV band.
More importantly, a sequence appeared: blazars with greater bolometric luminosity
had ``redder" SEDs (i.e. smaller peak frequencies: LBL in the terminology
of Padovani \& Giommi 1995), and the high energy peak
was more prominent.
Blazars of lower bolometric luminosity had instead a ``blue" SED
(HBL in the terminology of Padovani \& Giommi 1995) with the
two peaks having approximately the same luminosity. 

This spectral sequence was interpreted by Ghisellini et al. (1998) as due to 
the larger radiative cooling suffered by the emitting electrons of
blazars of larger power, where the radiation energy density
seen in their comoving frames received a large contribution by
photons produced outside the jet.
As a model, they adopted a simple leptonic one--zone synchrotron 
inverse Compton model.
In this scenario a stronger cooling resulted in a 
particle energy distribution with 
a break at lower energies (producing smaller peak frequencies) with more power 
emitted through the inverse Compton process (hence the dominance of the 
high energy peak).
This picture was later confirmed (Ghisellini, Celotti \& Costamante, 2002; 
Celotti \& Ghisellini 2008) when the new generations of Cherenkov 
telescopes allowed the detection of an increasing number of 
low power BL Lacs\footnote{see {\tt
http://www.mppmu.mpg.de/$\sim$rwagner/sources}}.
Through the modelling, the above studies found that there is a good correlation 
between the energy $\gamma_{\rm peak} m_e c^2$ of the electrons emitting at 
the peaks of the SED and the total (i.e. magnetic plus radiative) energy density 
as seen in the comoving frame of the blazar.
Therefore the so called ``blazar sequence" comes in two kinds:
i) a purely phenomenological sequence, relating the SED shape with the 
bolometric observed luminosity, and 
ii) a more ``theoretical" one, relating
$\gamma_{\rm peak}$ to the amount of radiative cooling.
It is clear that the first kind is associated with observed properties:
since it considers the brightest blazars, it is likely that it corresponds
to the most aligned sources: for blazars that are 
(even slightly) misaligned, the observed luminosity  of an intrinsically 
powerful source becomes smaller and the observed SED becomes
{\it ``redder"}, contrary to the general sequence.
Furthermore, the phenomenological blazar sequence had a rather
incomplete SED coverage (only 33 out of 126 sources were 
observed in the $\gamma$--ray band, at that time).
Thus, as it stands, the phenomenological blazar sequence probably 
describes active states of the sources, not necessarily
their averaged status.

The second kind, instead, treats comoving quantities and is less 
permeable to these effects. 
Since any reasonable blazar luminosity function predicts that the 
number of blazars increases for decreasing luminosities,
one simple prediction of the ``theoretical" scheme is that
there should be more ``blue" blazars than ``red" ones.

A critical review of the blazar sequence is presented by Padovani (2007),
who pointed out three main tests that the blazar sequence should pass:
\begin{enumerate}
\item 
existence of an (anti--)correlation between the synchrotron peak frequency
$\nu_{\rm s, peak}$ and the bolometric observed luminosity;

\item 
non--existence of ``blue" powerful objects;

\item
``blue" sources should be more numerous than ``red" ones.

\end{enumerate}
Point (i) concerns observed properties: since
red low luminosity blazars could be slightly misaligned sources, 
the existence of red blazar with small observed luminosities
is not invalidating the blazar sequence, that even
predicts them.
The other two points are more important.
In general, blazars with emission lines (FSRQs) have larger 
bolometric luminosities and ``red" spectra.
However, there are exceptions to this general rule.
One example is RGB J1629+4008 (at a redshift $z=0.272$), 
discussed in Padovani et al. (2002).
This blazar has broad emission lines with equivalent width 
typical of FSRQs ($\sim$80 \AA), and its synchrotron spectrum
peaks at $\sim$10$^{16}$ Hz, that corresponds to a ``blue" SED.
The luminosity of its broad emission lines
is relatively modest ($\sim$7$\times 10^{43}$ erg s$^{-1}$);
the modelling yields parameters consistent with the ``theoretical blazar scheme" 
(i.e. its $\gamma_{\rm peak}$ belongs to the correlation found in 
Ghisellini et al. 1998; see Fig. 9 of in Padovani et al. 2002).
Another object showing (weak) broad emission lines and a blue SED is
RX J1456.0+5048 ($z=0.479$)\footnote{
See the presentation by P. Giommi at\\ 
http://www.iasfbo.inaf.it/simbolx/program.php}.
As discussed in Maraschi et al. (2008) this source is very similar to 
RGB J1629+4008, and belongs as well to the $\gamma_{\rm peak}(U^\prime)$
correlation.

Padovani et al. (2003) searched for other blue quasars
using two large sample of blazars, the Deep X--ray Radio Blazar 
Survey (DXRBs), and the {\it ROSAT} All--Sky Survey--Green Bank 
Survey (RGB), for a total of about 500 sources.
They found a relatively large number of possible candidates,
but this finding was based on a rather poor characterisation
of the SED, parametrised through the radio, optical and 
X--ray fluxes and the broad band spectral indices connecting 
the radio to optical ($\alpha_{RO}$), the optical to X--ray 
($\alpha_{OX}$) and the radio to X--ray ($\alpha_{RX}$) fluxes.
Note that in the absence of multi--band optical data 
it is rather difficult to disentangle the beamed non--thermal
to the accretion disk blue bump continua.
For this reason these results were interesting,
but uncertain and therefore not conclusive.
A small fraction of these blue quasar candidates,
observed at the VLA (Landt, Perlman \& Padovani 2006), showed
a rather modest core radio luminosity, at the boundary
of the FR I and the FR II radio--galaxy division.
Of these, 10 sources were observed by {\it Chandra} 
(Landt et al. 2008) and showed {\it flat} X--ray spectra 
(i.e. energy spectral index $\alpha_X<1$), demonstrating that these 
blue quasars candidates have instead a red SED.
Also the Sedentary survey (e.g., Giommi et al. 2005), 
tuned to find high synchrotron peak sources, 
has not detected flat spectrum radio quasars.

Recently, two other high redshift and high power blazars were discovered,
claimed to be blue quasars candidates:
SDSS J081009.94+384757.0 ($z=3.95$; Giommi et al. 2007) and IGR J22517+2218 
($z=3.668$; Bassani et al. 2007). 
Both are characterised by a large X--ray to optical ratio, 
that could be interpreted as due to a single synchrotron 
component peaking at X--ray frequencies.
This possibility, mentioned in both papers,
is disfavoured by the fact that the X--ray spectral slope, 
in IGR J22517+2218,
is flatter than $\alpha_{OX}$, suggesting that 
the X--ray flux belongs to the high energy peak.
Instead, in SDSS J081009.94+384757.0, the optical flux 
can correspond to the thermal emission from the accretion 
disk, and if so, also this source fits very well in
the blazar sequence (see the SED and the corresponding model
in Fig. 3 of Maraschi et al. 2008).

% (by {\it UVOT},
% onboard Swift) describe a soft spectrum (i.e. $\alpha_0>1)$,
% while the X--ray spectrum (simultaneous, by {\it XRT} onboard Swift)
% is harder (even taking into account the
% relatively large error on the slope), and is inconsistent with being
% the extrapolation from the optical emission
% (see Fig. 3 in Maraschi et al. 2008).

The two papers mentioned above proposed also that if
these sources are indeed red, then they would be the
prototype of a new class of blazars with extreme properties,
namely a very dominating high energy peak.
However, these two sources have a SED that is
not unprecedented: the high redshift ($z>4$) blazars
(GB 1428+4217, $z = 4.72$:  Fabian et al. 1998;
PMN J0525--3343, $z=4.4$: Fabian et al. 2001;
RX J1028.6--0844, $z=4.276$: Yuan et al. 2000:
Q0906+6930,  $z=5.47$: Romani et al. 2004)
have a very similar SED, with a similar
X--ray to optical flux ratios (see also
RBS 315 at $z=2.69$; Tavecchio et al. 2007).
Therefore all these sources are not powerful blue quasars, 
but the extreme manifestation of the blazar sequence
(high energy peak increasingly dominating increasing the 
observed luminosity; see Maraschi et al. 2008).
Far from disproving the blazar sequence, they fully confirm it.

Concerning the third test that the blazar sequence
should pass (blue BL Lacs should be more numerous than red blazars), 
Padovani et al. (2007, see also the review of Padovani 2007)
pointed out that the ratio of blue/red counts of BL Lacs in deeper 
(in flux limit) radio and X--ray samples of blazars disagrees with the
predictions made by the Fossati et al. (1997) assuming
the blazar sequence.
The disagreement is moderate for the X--ray surveys, and is 
more severe for radio surveys down
to the 50 mJy flux level (in the sense that 
Padovani et al. 2007 find a ratio blue/red BL Lacs a factor $\sim$3
smaller than predicted by Fossati et al. 1997).

The possible solution to this problem offered by the blazar sequence 
is the following: if red blazars are intrinsically more powerful than 
blue BL Lacs, they can enter a flux limited sample even if the
jet is (slightly) misaligned, while blue BL Lacs cannot.
This selection effect could make red blazars to be over--represented in
the sample.
This is admittedly no more than an educated guess, and it should be supported
by detailed simulations: Fossati et al. (1997) did not consider
slightly misaligned blazars. 
As discussed below, the blazar sequence we are proposing in this paper
can offer an alternative (or additional) explanation to the above problem,
in terms of BL Lacs whose black hole has a relatively small mass.

In this paper we try to improve and to 
extend the blazar sequence, taking into account that

i) the phenomenological sequence was based on bright samples, 
and on a few high energy detections, not on the ``average" states
of typical sources;
ii) we now know many more TeV BL Lac objects;
iii) we made progress in calculating the power of the jet of blazars;
iv) we also made progress in estimating the mass of the central black hole
in some blazars.

The old phenomenological sequence was based on only one parameter:
the bolometric apparent luminosity.
We here explore the possibility to associate the SED of all blazars
to the two fundamental parameters of all AGNs: the mass of the black hole
$M$ and the accretion rate $\dot M_{\rm in}$.
To this aim we build upon the old ``theoretical" blazar sequence,
driven by a few simple ideas and pieces of evidence:

\begin{itemize}

\item There is a preferred region of the jet where most
of the observed radiation is emitted (see e.e. Ghisellini \& Madau 1996).
The location of this region, in the jet, should scale as
the mass of the black hole.

\item Broad emission lines come from a region whose distance
from the black hole scales (approximately, and with some scatter)
with the square root of the disk luminosity.

\item The broad emission line region exists only if the
disk luminosity is above a critical value (in Eddington units). 

\item The power of the jet (Poynting flux plus kinetic)
scales as the mass accretion rate.

\end{itemize}

We will examine and justify these points in the next section,
here we stress only that these main hypotheses will suffice to
completely describe the SED produced by the jets of blazars.
Therefore we will construct a two--parameter sequence,
for which the SED is dependent upon $\dot M_{\rm in}$ and $M$.
Since the disk luminosity $L_{\rm disk}$ is univocally determined 
given $\dot M_{\rm in}$ and $M$, the two parameters can equivalently 
be $L_{\rm disk}$ and $M$.

\section{Assumptions}
In this section we discuss the main assumptions of
our model.

\subsection{Black hole masses}

It is generally believed that the black hole mass of a radio--loud AGN
is on average larger than the typical mass of a radio--quiet object.
However, the exact range of black hole masses 
of radio--loud AGNs is still a matter of debate.

\begin{figure}  
\vskip -0.5 cm
\centerline{ \psfig{figure=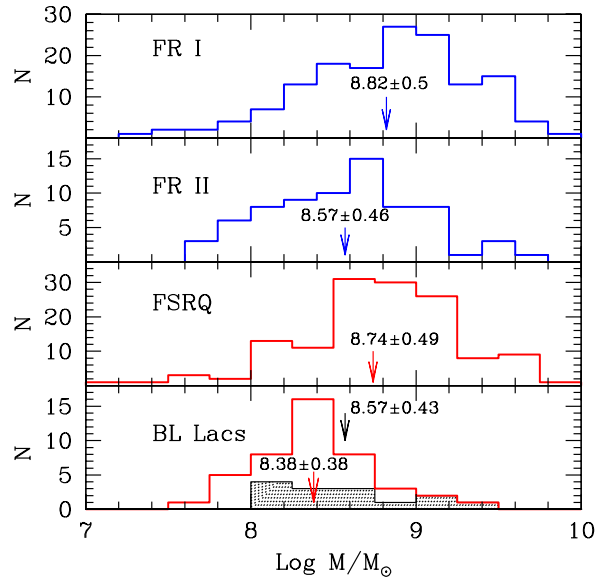,width=9cm} }
\vskip -0.5 cm
\caption{
Distribution of black hole masses calculated in Ghisellini \& Celotti (2001) 
for FR I and FR II radio--galaxies (top two panels)
by D'Elia et al. (2003) for FSRQs and for BL Lacs by Woo et al. (2005) and
Wagner (2008). In the bottom panel the shaded histogram is for TeV BL Lac only.
The arrows indicate the average of the logarithm of the masses, as labelled.
}
\label{mass}
\end{figure}

Lacy et al. (2001) proposed that the radio--loudness is a function of
mass, with most radio--loud sources having $M>$a few $\times 10^8$
$M_\odot$, while D'Elia et al (2003), with a sample of strong lined
blazars, found a range of masses extending significantly to the lower
end, down to a few $\times 10^7$ $M_\odot$ (see Metcalf \&
Magliocchetti 2006 for a recent discussion).  In Fig. \ref{mass} we
show the distributions of masses of FSRQs in the D'Elia et al. (2003)
sample and of BL Lac objects (Woo et al. 2005; Wagner 2008), and
compare them with the range of black hole masses estimated by
Ghisellini \& Celotti (2001) for FR I and FR II radio--galaxies.  A
caveat is in order: the correlations used to calculate the mass
(i.e. the mass--bulge optical luminosity, or the size of the broad
line region and the line width, or the $\sigma-M$ relation) have a
large scatter, and therefore the shown distribution are only
approximate.  We indicate in the figure the corresponding averages of
$\log M$ and their 1$\sigma$ dispersion.  
We conclude that most black hole masses are
in the range $3\times 10^7$--$3\times 10^9$ $M_\odot$, and use this
range in the following.

\subsection{Accretion rates}

A natural upper limit is obviously the Eddington one.  Another
critical accretion rate, probably connected with the change of
accretion mode, is for accretion luminosities $L_{\rm c}$ 
around a few $\times 10^{-3} L_{\rm Edd}$.  
Below this critical values the accretion flow
can be advection dominated (Narayan \& Yi 1994), or ADIOS (Blandford
\& Begelman, 1999, 2004), becoming radiatively inefficient, hotter and
geometrically thicker.  In Ghisellini \& Celotti (2001) we have
proposed that the dividing line between FR I and FR II (in a radio
luminosity vs host galaxy optical luminosity plot) corresponds to this
critical value.  There also are several evidences that BL Lac objects
indeed have radiatively inefficient disks, whose thermal emission is
never seen.  The absence of a strong ionising luminosity produced by
the disk explains in a natural way the absence (or the extreme
weakness) of the emission lines in BL Lacs.  Also, if this is the
reason of the FR I vs FR II and BL Lac vs FSRQ behaviour, there are
interesting consequences on the redshift evolution of these classes of
sources, since it is conceivable that there is some evolution in the
accretion rate (as in radio--quiet quasars), and therefore BL Lacs
(low accretion rates) might be more local (no or negative evolution),
while FSRQs should be more numerous (and powerful) in the past
(positive evolution), as proposed and discussed by Cavaliere \& D'Elia
(2002) and B\"ottcher \& Dermer (2002), and further discussed below.
As a lower limit to the accretion rate we are guided, for instance, by
M87, whose disk should emit at $L\sim 10^{-6} L_{\rm Edd}$.  Also with
these very small accretion rates, radio--loud systems can produce
powerful jets.

To summarise, we will assume that
for $L_{\rm c} / L_{\rm Edd} < L / L_{\rm Edd} < 1$ we have ``standard" accretion disks
which originates the jets in FSRQs (and FR II radio--galaxies),
while, for $10^{-6}< L/L_{\rm Edd}< L_{\rm c} / L_{\rm Edd}$ 
we have BL Lac objects (and FR I radio--galaxies).
The exact value  of this subdivision (i.e. $L_{\rm c}/L_{\rm Edd})$ 
is not very important, as long as the subdivision exists.

\subsection{Seed photons from the Broad Line Region}

For our purposes, one important parameter is the radiation
energy density of the broad emission line photons.
We therefore need to estimate the radius of the broad line
region and its total luminosity.
We assume that the latter is 10\% of the accretion disk luminosity
(if this is larger than $L_{\rm c}$, see above).
For the radius of the BLR there are, in the literature, several
proposals:
\begin{itemize}
\item 
according to Kaspi et al. (2005),
the relationship between the radius of the BLR $R_{\rm BLR}$ and
the ionising disk luminosity is:
\begin{equation}
{R_{\rm BLR} \over 10 {\rm ~~lt~~ days}} \, =\, (2.23\pm0.21) 
\left[ {\lambda L_\lambda (5100 {\rm \AA}) \over 10^{44} {\rm erg~~s^{-1}} }
\right]^{0.69\pm 0.05}
\label{r5100}
\end{equation}
where $L_\lambda (5100 {\rm \AA})$ is the monochromatic luminosity calculated at
5100 \AA.

\item
Bentz et al. (2006) pointed out a source of uncertainty
in this relation associated with the flux of the host galaxy, 
contributing more at low AGN luminosity.
Considering then a sample of AGN observed by HST, they derived:
\begin{equation}
\log R_{\rm BLR}  = (-21.2\pm 1.7) + (0.518\pm 0.04) 
\log \left[ \lambda L_\lambda (5100 {\rm \AA}) \right]
\label{r5100bis}
\end{equation}
with $R_{\rm BLR}$ in light days and $\lambda L_\lambda (5100 {\rm \AA})$
in erg s$^{-1}$.

\item
More recently, Kaspi et al. (2007) considered the CIV line and the 
continuum at 1350 \AA, deriving:
\begin{equation}
{R_{\rm BLR} \over 10 {\rm ~~lt~~ days}} \, =\, (0.17\pm 0.04) 
\left[ {\lambda L_\lambda (1350 {\rm \AA}) \over 10^{43} {\rm erg~~s^{-1}} }
\right]^{0.52\pm 0.04}
\label{r1350}
\end{equation}
Note that the last two relations have consistent slopes,
but inconsistent normalisations if $\lambda L_\lambda$
at 1350 and 5100 \AA\ is similar.

\end{itemize}

Given the above uncertainties, we chose to assume the simplest hypothesis,
which is a BLR radius scaling with the square root of the disk luminosity.
Also, for simplicity, we use the bolometric disk luminosity,
by assuming:
\begin{equation}
R_{\rm BLR} \, = \, 10^{17} L_{\rm disk,45}^{1/2} \,\,\, {\rm cm}
\label{rblr}
\end{equation}
This implies that the energy density of the line photons 
(for an observer at rest with the black hole) is constant:
\begin{equation}
U_{\rm BLR} \, =\, 0.1\, {L_{\rm disk} \over 4\pi R^2_{\rm BLR} c}
\, =\, 2.65\times 10^{-2}\,\,\, {\rm erg\, cm^{-3}}
\end{equation}
where we have also assumed a covering factor equal to 10\%.
Following Ghisellini \& Madau (1996), we then assume that the observer
comoving with the jet emission region measures $U^\prime_{\rm BLR}$ given by
\begin{equation}
U^\prime_{\rm BLR} \, =\, {17\over 12} \, \Gamma^2 \, U_{\rm BLR} \, =\,
3.76 \times 10^{-2} \Gamma^2 \,\,\, {\rm erg\, cm^{-3}}
\label{uext}
\end{equation}
The spectrum of this component is the sum of Doppler broadened lines
and a continuum.
The most prominent contribution comes from the Ly$\alpha$ line,
and this spectrum can be well approximated by a blackbody, 
with a peak (in $\nu F_\nu$, in the comoving frame) 
around $2\times 10^{15}\Gamma$ Hz
(see Tavecchio \& Ghisellini 2008 for a detailed discussion).

\subsection{Seed photons from the pc--scale dusty torus}

In the unification scenarios for Seyfert 1 and Seyfert 2 AGNs, one
assume the presence of a torus, at pc--scale distances, that blocks
the broad line photons to observers at large viewing angles, and
absorbs low energy X--ray photons, making the received X--ray spectrum
very hard.  Crucial to explain the X--ray background (Setti \& Woltjer
1989; Madau, Ghisellini \& Fabian 1994), the dusty torus could be
present also in jetted sources, but probably only in FR II
radio--galaxies (and FSRQ), since, in FR I radio--galaxies, the very
nucleus is not hidden in the optical (Chiaberge, Capetti \& Celotti
1999).  The possible importance of IR radiation from the torus for the
inverse Compton emission of blazars has been pointed out by
B{\l}a{\.z}ejowski et al. (2000) (see also Sikora et al. 2002, 2008).

The torus reprocesses the absorbed disk radiation into the IR band.
The typical temperature is around 150--200 K (Cleary et al. 2007),
as indicated by recent {\it Spitzer} observations.
We approximate the result of Cleary et al. (2007) by assuming that 
the torus reprocesses half of the disk radiation (corresponding
to a opening angle of $60^\circ$).
The typical distances of the torus, $R_{\rm IR}$, is assumed to scale 
as $L_{\rm disk}^{1/2}$, yielding a constant temperature 
($\propto L/R_{\rm IR}^2$). From the result of Cleary et al. (2007)
we then set:
\begin{equation}
R_{\rm IR} \, = \, 2.5\times 10^{18}  L_{\rm disk, 45}^{1/2}\,\,\, {\rm cm}
\end{equation}
The corresponding radiation energy density, as measured in the comoving frame, is
\begin{equation}
U^\prime_{\rm IR} \, =\, %{17\over 12} \, \Gamma^2 \, 
%{1\over 2} \, { L_{\rm disk}  \over 4\pi R^2_{\rm IR} c}\, =\,
3 \times 10^{-4} \Gamma^2 \,\, {\rm erg\, cm^{-3}}
\label{utorus}
\end{equation}
The spectrum of this component is assumed to be a blackbody,
with a peak in the comoving frame
(in a $\nu F_\nu$ plot) at $3\times 10^{13}\Gamma$ Hz.

\subsection{Location of the jet dissipation region}

We assume that the dissipation radius $R_{\rm diss}$ scales as the Schwarzschild
radius $R_s$:
$R_{\rm diss} = a R_s$, because this scaling is 
appropriate both in the case of 
an accelerating jet 
and in the scenario  of  internal shocks 
(Sikora, Begelman \& Rees 1994;
Ghisellini 1999; Spada et al. 2001) where we have
\begin{equation}
R_{\rm diss} \, \sim \, \Gamma^2 R_0 \, \sim a(\Gamma) R_{\rm s}
\label{rdiss}
\end{equation}
where $R_0$ is the initial separation of two consecutive shells, which
can be approximated as a multiple of the Schwarzschild radius.  We
allow ourselves to have a different factor $a$ for BL Lacs and for
FSRQs, to mimic the possibility to have two different origin for the
main dissipation in these two classes of objects: namely, in BL Lacs,
we could have a standing shock at some radial distance from the black
hole (along the lines of Sokolov, Marscher \& McHardy 2004).  Even in
this case we assume that this distance is proportional to the black
hole mass, but not necessarily with the same constant as in FSRQs.

The size of the emitting region is assumed to be a cylinder of cross
sectional radius $r= \psi R_{\rm diss}$ and height $\Delta R^\prime = r$,
as measured in the comoving frame. Here $\psi$ is the opening angle of
the jet.

\subsection{The jet power}

We will base our considerations upon the results 
presented in Celotti \& Ghisellini (2008).
In that work we found that the jet power $L_{\rm j}$ in blazars is
dominated by the protons associated to the emitting electrons.
We showed that the jet Poynting flux is smaller,
and also that the component of the jet power possibly transported
by electron positron pairs cannot be dynamically important
(see also Sikora \& Madejski 2000, Maraschi \& Tavecchio 2003).
In blazars the thermal component due to the accretion disk luminosity
is often hidden by the Doppler boosted non--thermal continuum of the jet,
but in FSRQs the presence of the broad emission lines allows to estimate
the disk luminosity.
We found that the power of jets
is comparable, and often larger than the luminosity
emitted by the accretion disk in FSRQs.
This is even more true in BL Lac objects, where the
disk radiation is almost always invisible.

Based on these considerations, we propose the following {\it ansatz}:
the jet power is always linked with the accretion rate, namely we can
write, both for BL Lacs and FSRQs:
\begin{equation}
L_{\rm j} \, = \, \eta_{\rm j}  \dot M_{\rm in} c^2 
\end{equation}
Since $L_{\rm disk} \, =\, \eta \dot M_{\rm in} c^2$,
the jet power can always be written as
\begin{equation}
L_{\rm j} \, = \, {\eta_{\rm j} \over \eta} \, L_{\rm disk}
\label{lk}
\end{equation}
If we set, in FSRQs, the jet efficiency factor $\eta_{\rm j}$
equal to the disk accretion efficiency $\eta$ (in the standard accretion mode,
i.e. $\eta_{\rm j}=\eta \sim 0.1$), we have $L_{\rm j} = L_{\rm disk}$.  
If, as it seems the case (Celotti \& Ghisellini 2008; Maraschi \& Tavecchio
2003) the jet power exceeds the disk luminosity, then $\eta_{\rm j}>\eta$.

We assume that the accretion rate in BL Lac objects
is below a critical value,  $\dot M_{\rm c}$, at
which the accretion flow changes regime from
``standard" disk accretion to ADAF--like or ADIOS--like regimes,
for which
%       $L_{\rm disk} \propto \dot M_{\rm in}$ for $\dot M_{\rm in} > \dot M_{\rm c}$
%
$L_{\rm disk} \propto \dot M_{\rm in}^2 $
(e.g., Narayan, Garcia \& McClintock 1997).

We must then relate $\dot M_{\rm in}$ with the disk luminosity and
then relate the disk luminosity with the jet kinetic power.  Since at
$\dot M_{\rm c}$ the two accretion regimes corresponds to the same
$L_{\rm disk}=L_{\rm c}$, we have 
\begin{eqnarray}
\eta &=& 0.1; \qquad  
\,\,\,\, {\dot M_{\rm in} \ge \dot M_{\rm c}
\,\, {\rm or}\, \, L_{\rm disk}\ge L_{\rm c}}
\nonumber \\
\eta &=& 0.1 \, {\dot M_{\rm in}\over \dot M_{\rm c}};
\,\, \,\,\,  {\dot M_{\rm in} < \dot M_{\rm c}
\,\, {\rm or}\,\, L_{\rm disk}\le L_{\rm c}}
\label{dotm}
\end{eqnarray}
where we set $L_{\rm c} =  \eta \dot M_{\rm c} c^2$.

The above assumptions allows to consider $L_{\rm disk}$ (or $L_{\rm j}$)
as the other key parameter, beside the black hole mass, instead 
of $\dot M_{\rm in}$. 
The obvious advantage is that both $L_{\rm disk}$ and $L_{\rm j}$
can be derived (or even directly observed, in the case of $L_{\rm disk}$
of FSRQs) in a much easier way than $\dot M_{\rm in}$.

The above {\it ansatz} implies that the outflowing mass rate can
be found through
\begin{equation}
L_{\rm j} \, =  \Gamma \dot M_{\rm out} c^2\,
\to \, \dot M_{\rm out} \, =\, {\eta_{\rm j} \over  \Gamma} \dot M_{\rm in}
\end{equation}

\subsection{Jet Poynting flux and kinetic power} 

We assume that a fraction $\epsilon_B$ 
of $L_{\rm j}$ is in the form of Poynting flux.
This allows us to estimate the magnetic energy density as:
\begin{equation}
L_{\rm B} = \epsilon_{\rm B} L_{\rm j} =
\pi \psi^2 R_{\rm diss}^2 U_B \Gamma^2 c  \to 
U_{\rm B}  = {\epsilon_{\rm B} L_{\rm j} \over \pi \psi^2 R_{\rm diss}^2 \Gamma^2 c}
\label{ub}
\end{equation}
We assume that a fraction $\epsilon_{\rm e}$ is converted, 
at $R_{\rm diss}$, in relativistic electrons. 
Their energy density, calculated in the comoving frame,
is $U_{\rm e} \equiv m_{\rm e} c^2 \int N(\gamma) d \gamma$ and is given by
\begin{equation}
L_{\rm e} = \epsilon_{\rm e} L_{\rm j} \, \to \, 
U_{\rm e}  = {\epsilon_{\rm e} L_{\rm j} \over \pi \psi^2 R_{\rm diss}^2 \Gamma^2 c}
\label{ue}
\end{equation}
In fast cooling (i.e. when all the injected electrons can cool in one
light crossing time) all the power in electrons is converted into radiation.
In slow cooling, instead, only a fraction $\epsilon_{\rm A}$ can be 
emitted. 
Therefore we have that the energy density of the radiation produced
by the jet is
\begin{equation}
U_{\rm r}  \, =\,  \epsilon_{\rm A} U_{\rm e} \, =\,
{\epsilon_{\rm e} \epsilon_{\rm A} L_{\rm j} \over 
\pi \psi^2 R_{\rm diss}^2 \Gamma^2 c}
\end{equation}
In our scenario $\epsilon_{\rm e}$ and $\epsilon_B$ are free parameters,
while $\epsilon_{\rm A}$ can be derived, as described in the Appendix.
In Celotti \& Ghisellini (2008) $\epsilon_B$ and $\epsilon_{\rm e}$
have been found (through spectral modelling) 
for a sample of BL Lacs and FSRQs detected in $\gamma$--rays.
It was found that there exists a good correlation between $\epsilon_{\rm e}$
and $L_{\rm j}$, and a much more scattered correlation (or trend)
between $\epsilon_B$ and $L_{\rm j}$.
Using the relations found in that paper, we here set:
\begin{equation}
\epsilon_{\rm e} \, \sim\,  0.1\, L^{-0.4}_{\rm j, 45} 
\label{epse}
\end{equation}
\begin{equation}
\epsilon_B\, \sim\, 0.05\,  L^{-0.25}_{\rm j, 45}  
\label{epsb}
\end{equation}

\subsection{Energy of the electrons emitting at the peaks of the SED} 

Let us call $\gamma_{\rm peak}$ the random Lorentz factor of the
electrons emitting at the two peaks of the SED.
We take advantage of the 
observed correlation between $\gamma_{\rm peak}$ and the (comoving)
energy density $U^\prime(\gamma_{\rm peak})$ that is the sum
of the magnetic and radiation energy density integrated up to
$\nu = m_{\rm e} c^2/(h\gamma_{\rm peak})$.
In other words, $U^\prime(\gamma_{\rm peak})$ is the sum of the magnetic 
and the radiation energy density available
for scattering in the Thomson regime (i.e. efficient cooling).
This correlation (updated in Celotti \& Ghisellini 2008) is of the form
\begin{eqnarray}
\gamma_{\rm peak} \,&=&\, 10^3 \, \left( {U'\over 0.3}\right)^{-1};
\qquad U^\prime \le 0.3 {\rm ~~erg~~cm^{-3}} \nonumber \\
\gamma_{\rm peak} \,&=&\, 10^3 \, \left( {U'\over 0.3}\right)^{-1/2};
\quad U' \ge 0.3 {\rm ~~erg~~cm^{-3}}  
\label{gpeak} 
\end{eqnarray}
In Ghisellini et al. (2002) we have interpreted this correlation
(found by modelling SEDs of blue BL Lac objects) as the
result of the radiative cooling occurring in one 
light crossing time. 
In this time electrons cool down to an energy $\gamma_{\rm c}$ given by
\begin{equation}
\gamma_{\rm c} \, =\, {3 m_e c^2 \over 
4\sigma_{\rm T} \Delta R^\prime U'} \, \propto \, U'^{-1}
\label{gcool} 
\end{equation}
This scaling stops when $\gamma_{\rm c}<\gamma_1$, because in this case
the peak (in $\nu F_\nu$) is produced by electrons at $\gamma_1$.
According to Celotti \& Ghisellini (2008), this occurs when $\gamma_{\rm c} \sim 10^3$.
For larger energy densities, the scaling $\gamma_{\rm peak} \propto U'^{-1/2}$
suggests that it is the cooling rate (at $\gamma_{\rm peak}$) which is constant, 
i.e. $\dot\gamma \propto \gamma_{\rm peak}^2U'$ is the same for all powerful blazars.
Note, however, that the scatter in the
region of large $U'$ is large, making this relation true 
only approximately.

\subsection{The emitting particle distribution}

The energy distribution of the emitting particles, $N(\gamma)$ [cm$^{-3}$], 
is assumed to be the same as the one described in 
Ghisellini, Celotti \& Costamante (2002) and Celotti \& Ghisellini (2008).
This distribution approximates the case of
an injection of particles, throughout the source,
lasting for a finite injection time, equal to 
$t_{\rm inj}=\Delta R^\prime /c = \psi R_{\rm diss}/c$.
This is because blazars are variable (flaring) sources, 
and a reasonably good representation of the observed spectrum 
can be obtained by considering the particle distribution at the 
end of the injection.
When the injection stops, particles above $\gamma_{\rm c}$ have cooled,
modifying the energy distribution of the injected particles.
The latter is assumed to be a broken power--law with slopes $\propto
\gamma^{-1}$ and $\propto \gamma^{-s}$ below and above the break at
$\gamma_{\rm inj}$.  
We assume that $2<s<3$.
In the case of fast cooling, occurring when $\gamma_{\rm c} < \gamma_{\rm inj}$,
we have:
\begin{eqnarray}
N(\gamma) & \propto &\gamma^{-(s+1)}; \qquad \gamma > \gamma_{\rm inj}
\nonumber\\
N(\gamma) & \propto &\gamma^{-2}; \qquad \quad\,\,\,\, 
\gamma_{\rm c} < \gamma < \gamma_{\rm inj}
\nonumber\\
N(\gamma) & \propto &\gamma^{-1}; \qquad \quad\,\,\,\,  \gamma < \gamma_{\rm c}
\end{eqnarray}
Since $s>2$, in this case the peak energy $\gamma_{\rm peak}=\gamma_{\rm inj}$.
When instead $\gamma_{\rm c} > \gamma_{\rm inj}$ (slow cooling), we have:
\begin{eqnarray}
N(\gamma) &\propto& \gamma^{-(s+1)}; \qquad \,\,\, 
\gamma> \gamma_{\rm c} \nonumber\\
N(\gamma) &\propto& \gamma^{-s};\qquad \qquad\gamma_{\rm inj} <\gamma <
\gamma_{\rm c}\nonumber\\
N(\gamma) &\propto& \gamma^{-1};\qquad \qquad\gamma < \gamma_{\rm inj}.
\end{eqnarray}
For the assumed range of values of $s$, the peak energy in this case is $\gamma_{\rm peak}=\gamma_{\rm c}$.

\subsection{Summary of free parameters}

Our scheme needs the following parameters:

\begin{itemize}
\item the black hole mass $M$;
\item the accretion rate $\dot M_{\rm in}$, or, equivalently, the 
disk luminosity in units of the Eddington one, $L_{\rm disk}/L_{\rm Edd}$;
\item the bulk Lorentz factor $\Gamma$;
\item the initial separations of the colliding shells (in the internal shock case)
or, more generally, the distance, in units of Schwarzschild radii, of the dissipation region;
\item the ``equipartition" parameters $\epsilon_{\rm e}$ and $\epsilon_B$
\item the jet efficiency factor $\eta_{\rm jet}$;
\item the slope $s$ of the injected particle distribution.
\end{itemize}
The viewing angle $\theta$ is not considered as a free parameter, since
we always assume $\Gamma\sim\delta$ and then $\theta\sim 1/\Gamma$.
We have rather good observational constraints on several of these 9 parameters.
For instance, $10<\Gamma<20$; 
$R_{\rm diss}$ is a few hundreds Schwarzschild radii; 
the (average) slope of the injected particle distribution is $s\sim$ 2.4--2.5; 
the jet is slightly more powerful than the disk luminosity,
leading to $\eta_{\rm jet}\sim$ 0.2--0.5; 
as mentioned, we have derived $\epsilon_B$ and $\epsilon_{\rm e}$ for a number of 
blazars, and we can use these numbers in our scenario.
In general, the most important parameters are thus $M$ and $\dot M_{\rm in}$,
that are the leading quantities characterising our proposed scheme.

\begin{figure}  
\vskip -0.5 cm
\centerline{
\psfig{figure=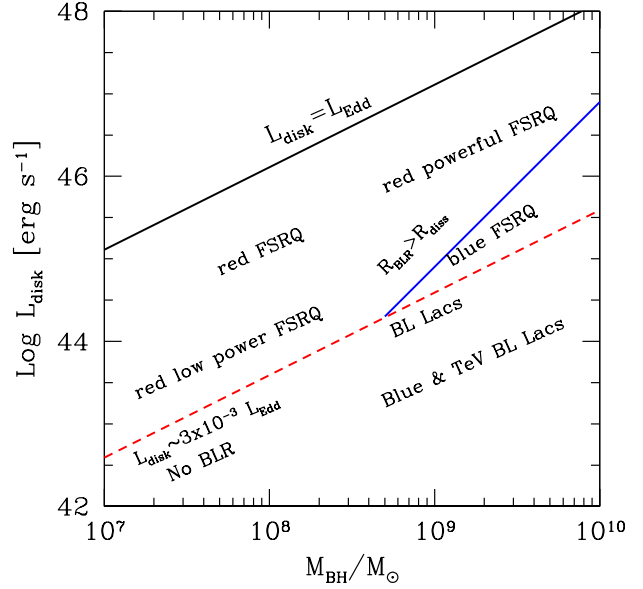,width=9cm}}
\vskip -0.5 cm
\caption{
The disk luminosity vs black hole mass plane.
Different blazar classes occupy different regions of this plane.
The dashed line corresponds to a disk luminosity equal to
a fraction of the Eddington one (for this figure we have used
$3\times 10^{-3}$).
Below this line we assume that the broad line region does not
exist, or it is very weak. Therefore BL Lac objects live
below this line: ``classic" BL Lac objects just below, and
TeV BL Lacs at smaller still disk luminosities.
According to our scenario, there is another important dividing line,
corresponding to where $R_{\rm diss}> R_{\rm BLR}$
[for this figure we have used $a(\Gamma)=300$].
Objects having a BLR, but whose jets preferentially dissipates beyond it,
are ``blue" quasars.
If blazars can have relatively small black hole masses (i.e.
even smaller than $10^8$ $M_\odot$), then there should be a population
of relatively low power quasars, with a red spectrum.
}
\label{plane}
\end{figure}

\section{Simple consequences}

In this section we derive in an heuristic way a few simple
consequences of our assumptions, leaving for the following sections
a more detailed description, which requires more technical details.
\begin{itemize}
\item 
{\bf BL Lac/FSRQ division ---} The first consideration is, in itself, 
one of our assumptions: the BLR exists
only when $L_{\rm disk}/L_{\rm Edd}$ is greater than some value, which we take
equal to $3\times 10^{-3}$.
This immediately implies a division between FSRQs and BL Lac objects,
defined as objects with and without broad emission lines, respectively.

\item 
{\bf Existence of ``blue" FSRQs ---} The BLR radius $R_{\rm
BLR}\propto L_{\rm disk}^{1/2} \propto \dot M_{\rm in}^{1/2}$, while
the dissipation distance is $R_{\rm diss} \propto M$.  Therefore there
is the possibility that relatively high mass objects, with relatively
faint disk (but with $L_{\rm disk} > L_{\rm c}$) have jets
preferentially dissipating beyond the BLR (see also Georganopoulos,
Kirk \& Mastichiadis 2001, Pian et al. 2006). The emitting electrons
would suffer less radiative cooling, implying a large $\gamma_{\rm
peak}$ and a ``blue" SED. From Eq. \ref{rdiss} and Eq. \ref{rblr} we
have that $R_{\rm diss} > R_{\rm BLR}$ for

\begin{equation}
0.39 M_9 < L_{\rm disk, 45} \, <\, 0.81 M_9^2 
\left[ { a(\Gamma) \over 300} \right]^2
\end{equation}
This defines a ``triangular" region in the $L_{\rm disk}$--$M$ plane,
as illustrated in Fig. \ref{plane}.

\item 
{\bf Existence of ``red" FSRQs at relatively low power ---}
This is a consequence of considering relatively small black hole masses.
If FSRQs with --say-- $M\sim 3\times 10^7$ $M_\odot$ exist, then they 
have disk luminosities down to $L_{\rm disk} \sim 1.2\times 10^{43}$ erg s$^{-1}$;
and similar jet powers (see Fig. 2).
Dissipation takes place well within the (small) BLR, 
in regions of large $U^\prime$ implying small $\gamma_{\rm peak}$
and hence a red SED.
The emitting regions would also have 
relatively large magnetic fields ($U_B\propto 1/M^2$, see Eq. \ref{ub}),
implying large $U_B$.
The synchrotron self--Compton emission would then be relatively more
important than the External Compton component.

\item
{\bf The Compton dominance ---}
It is defined as the ratio of the inverse Compton to synchrotron luminosity
$L_{\rm C}/L_{\rm s}$ and corresponds to (for scattering in the Thomson regime)
\begin{equation}
{ L_{\rm C} \over L_{\rm s} }\, \sim \, 
{U^\prime_{\rm s} +U^\prime_{\rm ext} \over U_B}
\end{equation}
where the external radiation energy density $U^\prime_{\rm ext}$
(in the comoving frame) is produced by the BLR and by the dusty torus, 
while $U^\prime_{\rm s}$ is the synchrotron radiation energy density.
In general, we expect that the Compton dominance is larger for
FSRQs with $R_{\rm diss} < R_{\rm BLR}$.
But since $U_B$ is larger for smaller $R_{\rm diss}$ (smaller masses),
while $U^\prime_{\rm ext}$ is constant (for equal $\Gamma$),
we expect that the Compton dominance is on average larger for larger masses.

\end{itemize}

In Fig. \ref{plane} we illustrate these simple points in the plane 
$L_{\rm disk}$--$M$.
The line dividing BL Lac objects from FSRQ is assumed to
to be at $L_{\rm disk}/L_{\rm Edd}=3\times 10^{-3}$, 
the radius of the broad line
region corresponds to Eq. \ref{rblr}, and the dissipation radius
assumes $a(\Gamma)=300$.
These values are indicative, but not certain, and  there can be
a substantial scatter.
We recall that ``red" and ``blue" 
here mean that the synchrotron and inverse Compton
peak frequencies have low or large values, respectively, but 
not that they are necessarily more or less Compton dominated.
To make further predictions about the predicted SED as a function
of $M$ and $\dot M_{\rm in}$, we must evaluate the Compton
dominance in detail. This is done below.

\section{Estimating the Compton dominance}

In this section we show how it is possible, with a few reasonable assumptions,
to derive the ratio between the inverse Compton to synchrotron luminosity. 
This parameter, called Compton dominance, will determine in what region of 
the $M$--$\dot M_{\rm in}$ plane we can find  $\gamma$--ray 
bright sources (and viceversa where are the $\gamma$--ray dim blazars).

In the Appendix we show the importance of the so--called Comptonization
$y$ parameter, and how it can be calculated. 
We define it as
\begin{equation}
y\, \equiv \, \sigma_{\rm T} \Delta R^\prime  \int \gamma^2 N(\gamma) d\gamma
\label{def_y}
\end{equation}
In the Appendix we also show that the comoving  synchrotron 
and self--Compton radiation energy densities
can be expressed as 
$U^\prime_{\rm s} = y U_B$ and $U^\prime_{\rm SSC} = y^2 U_B$, 
while the external Compton energy density is $U^\prime_{EC} = y U^\prime_{\rm ext}$.
The asymptotic values of $y$ are:

\begin{eqnarray}
y\, &\sim & 
{1\over 2 } \left[ 1+ {4\epsilon_{\rm e}\epsilon_{\rm A} \over  
\epsilon_{\rm B} } \right]^{1/2}-{1\over 2}; \, 
\qquad { U^\prime_{\rm ext}\over U_{\rm B} }  \ll 1  \nonumber \\
y\, &\sim & \, { \epsilon_{\rm e}\epsilon_{\rm A}\over \epsilon_{\rm B} } \, 
{ U_{\rm B} \over U^\prime_{\rm ext} };
\qquad \qquad \qquad \quad \,\,  { U^\prime_{\rm ext}\over U_{\rm B} } \gg 1
\label{y_array}
\end{eqnarray}
where we have assumed that the scattering process occurs in the Thomson
regime. 
In this regime the ratio between the inverse Compton and the
synchrotron luminosities can be approximated as
\begin{equation}
{L_{\rm c}\over L_{\rm s} } \, =\, {U'_{\rm syn}\over U_{\rm B}}+
{U'_{\rm ext} \over U_{\rm B}}\, =\, y + {U'_{\rm ext} \over U_{\rm B}}
\label{urub} 
\end{equation}
The first term on the RHS controls the relative importance of 
the SSC component, while the second measures the relative
importance of the EC component.
They are not independent, because the value of $y$ is in turn
controlled by the $(U'_{\rm ext} / U_{\rm B})$ ratio, as shown
by Eq. \ref{y_array}.

% ================================================================== 
% ================================================================== 
\begin{figure}
\vskip -1 true cm
% \centerline{ 
\hskip -2.5 true cm
\psfig{figure=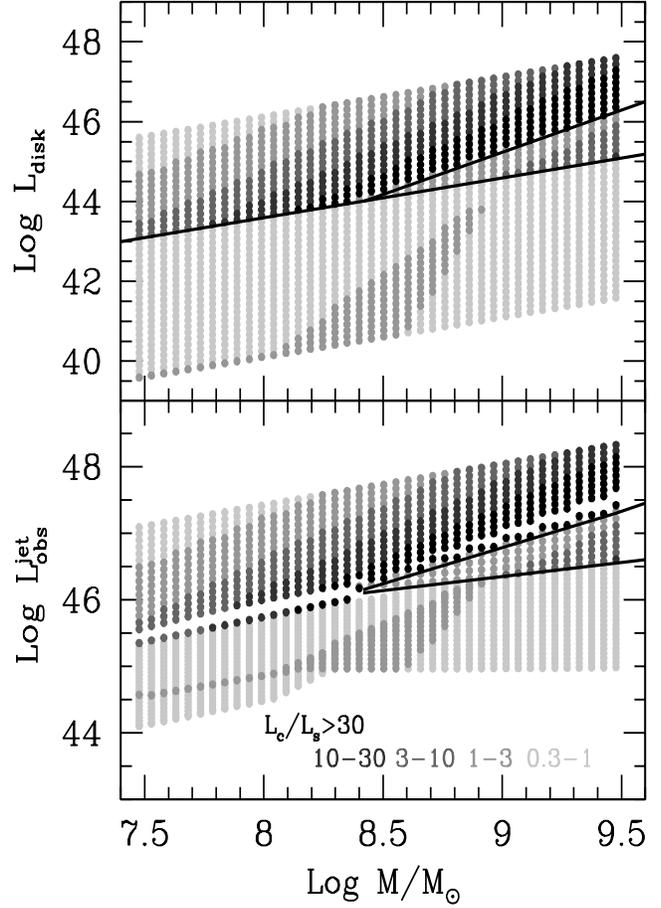,width=12.7cm,height=14.5cm} % }
\vskip -1 true cm
\caption{
The disk accretion luminosity (top panel) and the observed 
bolometric jet emission (bottom panel), as a function of black hole mass.
The bulk Lorentz factor is assumed to be 15 for all cases,
while we have assumed $a(\Gamma)=300$ for FSRQs and 100 for BL Lacs.
The solid lines on bot panels correspond to 
$L/L_{\rm Edd}=3\times 10^{-3}$ and on to the region
where $R_{\rm diss}>R_{\rm BLR}$.
The different grey levels correspond to different ``Compton dominance",
i.e. different $L_{\rm C}/L_{\rm s}$ ratios.
}
\label{ml}
\end{figure}

\begin{figure}
\vskip -1 true cm
\hskip -2.5 true cm
\psfig{figure=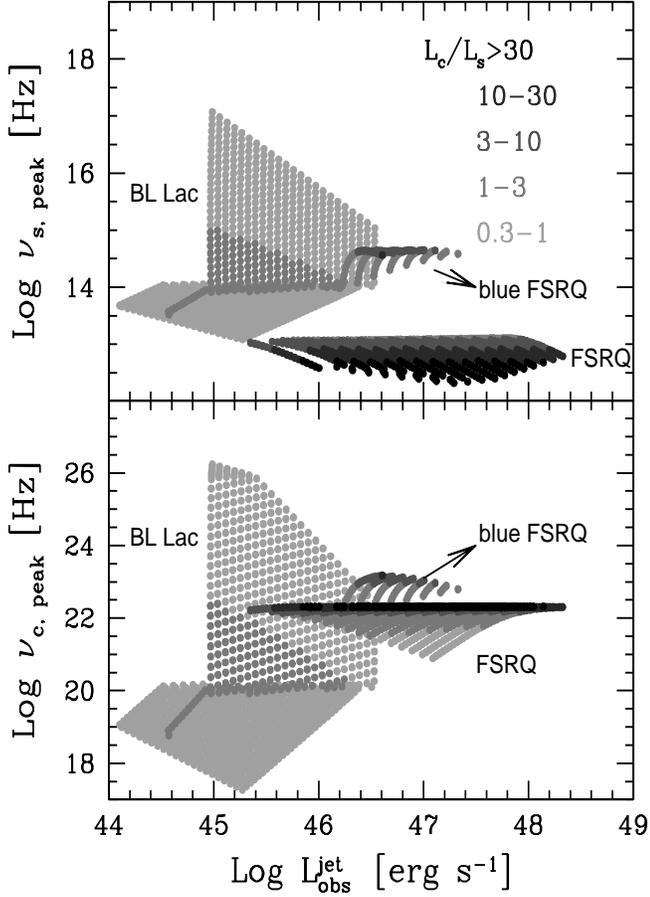,width=12.7cm,height=14.5cm} 
\vskip -1 true cm
\caption{
The peak synchrotron frequency $\nu_{\rm s, peak}$ 
(top) and  the peak Compton frequency $\nu_{\rm c, peak}$ (bottom)
as a function of the observed jet luminosity. 
As in Fig. \ref{ml}, different grey levels indicate
different Compton dominance values.
}
\label{vp}
\end{figure} 
% ================================================================== 

Isolating the EC dominance we can focus on FSRQs with
$R_{\rm diss}< R_{\rm BLR}$:

\begin{equation}
{L_{\rm EC}\over L_{\rm s} } \, =\,  {U'_{\rm ext} \over U_{\rm B}} 
\propto {M^2 \over 
\dot M_{\rm in}} \Gamma^4 { [a(\Gamma)]^2 \over\epsilon_{\rm B}}
\propto
{ M \over  L_{\rm j}/L_{\rm Edd}} \Gamma^4 
{[a(\Gamma)]^2 \over\epsilon_{\rm B}}
\label{lcls} 
\end{equation}
where we made use of Eq. \ref{uext}, Eq. \ref{rdiss} and Eq. \ref{ub}.
This implies that FSRQs of 
the same mass and increasing $\dot M_{\rm in}$
form a sequence of increasing power and {\it decreasing}
Compton dominance.
On the other hand FSRQs of increasing mass and constant 
$L_{\rm j}/L_{\rm Edd}$  
form a sequence of increasing power {\it and} increasing Compton
dominance. 
Note also the strong dependence on the bulk Lorentz factor.
FSRQs of equal masses and accretion rates can form a sequence of
increasing power and increasing Compton dominance by having
slightly different values of $\Gamma$.

The SSC over synchrotron power ratio is simply $y$.
When the EC emission is unimportant, 
$y\sim (\epsilon_{\rm e}\epsilon_{\rm A}/ \epsilon_{\rm B})^{1/2}$, 
greater than unity when 
$\epsilon_{\rm e} > \epsilon_{\rm B}/\epsilon_{\rm A} $
(Eq. \ref{y_array}).

\section{Results}

The results of our scheme, for the set of chosen parameters,
are shown in Fig. \ref{ml} and Fig. \ref{vp}.
In this figures the different grey levels corresponds to different
Compton dominance values, as labelled.
Consider first Fig. \ref{ml}.
The top panel is equivalent to Fig. \ref{plane}.
Below the diagonal line $L_{\rm disk}/L_{\rm Edd} =3\times 10^{-3}$ 
we have BL Lacs, above the line we have FSRQs.

The solid lines divide the plane in three regions:
in region I, above the solid line
$L_{\rm disk}/L_{\rm Edd} =3\times 10^{-3}$,
we have the FSRQs.
Below this line, in region II, we have BL Lac objects.
In region III, limited by the two solid lines,
we have the ``blue" FSRQs.

The Compton dominance in region I increases for increasing mass
and decreases for increasing $L_{\rm disk}$ for objects with the same mass,
as described by Eq. \ref{lcls} [for $a(\Gamma)$ constant].
In region II the Compton dominance, determined by 
$y\sim (\epsilon_{\rm e}\epsilon_{\rm A}/ \epsilon_{\rm B})^{1/2}$, 
is varying much less and is of order unity in the entire region.
In region III we have the same trend as in region I, but
with a lower Compton dominance value, determined partly 
by SSC emission (scaling with $y$) and partly by the
torus external IR photons.

The bottom panel shows the observed bolometric luminosity 
produced by the jet as a function of the mass.
In this plane objects belonging to regions II and III partially overlap,
since we can have ``blue" FSRQs slightly less
luminous than BL Lacs. This effect is due
to the fact that in this illustrative figure we have assumed that
the dissipation region of BL Lac objects has $a(\Gamma)=100$,
while we use $a(\Gamma)=300$ for FSRQs.
Some BL Lac objects, therefore, although lacking the BLR, 
are slightly more efficient (larger $\epsilon_{\rm A}$)
than ``blue" FSRQs, simply because they are more compact.
This also means that their magnetic field is relatively larger,
resulting in a less Compton dominated source.

In Fig. \ref{vp} we show the predicted peak frequency of the synchrotron
flux (top panel) and of the IC emission (bottom).
Again, different grey levels correspond to different Compton dominance
values.
Note that the range of $\nu_{\rm s, peak}$ and $\nu_{\rm c, peak}$ 
values is much broader for BL Lac objects than for FSRQs.
This is the consequence of two facts: first, the dependence
of $\gamma_{\rm peak}$ on $U^\prime$ is stronger for BL Lacs
than for FSRQs
($\gamma_{\rm peak} \propto U'^{-1}$ vs  
$\gamma_{\rm peak} \propto U'^{-1/2}$ respectively);
secondly, the energy density of the external seed photons 
is constant, as implied by assuming 
$R_{\rm BLR}\propto L_{\rm disk}^{1/2}$ and a constant $\Gamma$: thus
when the external radiation energy density dominates, it 
implies a quasi--constant $\gamma_{\rm peak}$,
that in turn makes $\nu_{\rm s, peak}$ and $\nu_{\rm c, peak}$ 
to vary in a small range.

Fig. \ref{vp} shows that the maximum values of $\nu_{\rm s, peak}$ 
and $\nu_{\rm c, peak}$ of FSRQs have a sharp boundary.
This is due to the following.
When the BLR radiation energy density dominates the cooling,
we have
\begin{equation}
\nu_{\rm s, peak} \, =\, 3.6\times 10^6 \Gamma B \gamma_{\rm peak}^2\,
=\, { 3.6\times 10^6  \Gamma B \over (17/12)U_{\rm BLR} \Gamma^2 +U_B  }
\, =\, {2.7\times 10^{13} \Gamma B \over 0.944 \Gamma^2 +B^2}\, \, \, {\rm Hz}
\label{vsmax}
\end{equation}
where we have used Eq. \ref{uext} and Eq. \ref{gpeak}.
The maximum $\nu_{\rm s, peak}$ is reached for $B^2 = 0.944 \Gamma^2$.
In our case, since $\Gamma=15$, we have $\nu_{\rm s, peak}=1.4\times 10^{13}$ Hz.

Blue FSRQs, on the other hand, suffer much less cooling (they dissipate beyond the BLR)
and consequently their $\gamma_{\rm peak}$ and $\nu_{\rm s, peak}$ 
are larger, explaining the existence of the gap (in $\nu_{\rm s, peak}$)
of the two populations.

It is worth to stress that blazars will not all have the same value of $\Gamma$
(that can change even in a single source).
The sharp boundaries described above and shown in Fig. \ref{vp} will disappear once we allow 
for a distribution of $\Gamma$--values.

% -------------------------------------------------------
\begin{figure}
\vskip -0.75 true cm
\hskip -1.8 true cm
\psfig{figure=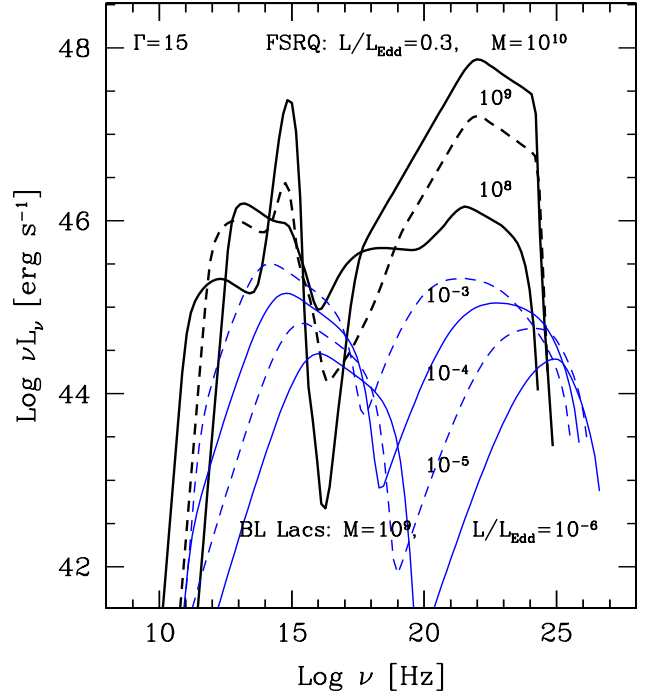,width=11cm}  
\vskip -0.8 true cm
\caption{
Predicted SED as a function of black hole mass and disk luminosities.
For FSRQ (top three thick lines) we have assumed a constant $L/L_{\rm Edd}$ ratio
(equal to to 0.3) and different black hole masses.
For BL Lacs, instead (lower four lines thin lines), we the shown SEDs correspond to 
a black hole mass of $10^9$ $M_\odot$ and different $L/L_{\rm Edd}$ values, 
as labelled.
}
\label{sedgamma15}
\end{figure} 
% -------------------------------------------------------

\subsection{Predicted SED}

In Fig. \ref{sedgamma15} we show several predicted
SEDs of FSRQs and BL Lac objects.
For FSRQs the spectra correspond to objects 
with the same ratio $L_{\rm disk}/L_{\rm Edd}=0.3$ and different masses,
while for BL Lacs we have the same mass ($10^9$ $M_\odot$)
and different $L_{\rm disk}/L_{\rm Edd}$ ratios.
In agreement with Fig. \ref{ml}, FSRQs are more Compton dominated
as the mass (and then the bolometric luminosity, for constant
$L_{\rm disk}/L_{\rm Edd}$) increases, with almost constant peak frequencies.
Note that, increasing the mass, the IC luminosity increases,
while the synchrotron component decreases.
This can be understood recalling that
$L^\prime_{\rm s} \propto y U_B R_{\rm diss}^2$
and inserting the appropriate scaling of $U_B$ (Eq. \ref{ub}) and
$y$ (Eq. \ref{ysolved}), yielding
$L^\prime_{\rm s} \propto \epsilon_{\rm e}\epsilon_B L_{\rm j}^2/R_{\rm diss}^2
\propto \epsilon_{\rm e}\epsilon_B (L_{\rm disk}/L_{\rm Edd})^2$.
Therefore, for a constant ratio $L_{\rm disk}/L_{\rm Edd}$,
the synchrotron luminosity $L_{\rm s} \propto \epsilon_{\rm e}\epsilon_B$
which we assume to scale as $L_{\rm j}^{-(0.4+0.25)}$
(see Eq. \ref{epse} and Eq. \ref{epsb}).

The spectral indices in the $\sim$ GeV ($\alpha_\gamma$) 
and X--ray ($\alpha_X$) bands
anti--correlate: at low luminosities we have a steep $\alpha_X$ (due to the
tail of the synchrotron emission) and a hard $\alpha_\gamma$ (due to 
a still rising SSC spectrum), while at large powers 
we have a flat $\alpha_X$ (due to the low energy part of the EC spectrum)
and a steep $\alpha_\gamma$ (due to the high energy tail of the same
component).
This is in agreement to what observed (e.g. Comastri et al. 1997).

The series of SED in Fig. \ref{sedgamma15} resembles the
phenomenological sequence of Fossati et al. (1998). 
In order to
reproduce the observed increase of the Compton dominance with the
total power for FSRQs we assume the same $L_{\rm disk}/L_{\rm Edd}$
and change the black hole mass. This choice is dictated by the fact
that, in the present scheme, the Compton dominance increases with the
mass (as clearly visible in Fig. \ref{ml}). 
Fixing the mass and
increasing the disk luminosity, instead, would have the effect to {\it
reduce} the Compton dominance, contrary to the observed trend.  
Low power BL Lac objects, instead, display in the blazar sequence a more
or less constant Compton dominance and we can reproduce this branch of
the blazar sequence using a fixed (large) black hole mass and varying
the accretion rate.

\begin{figure}
\vskip -0.8 true cm
\centerline{
\psfig{figure=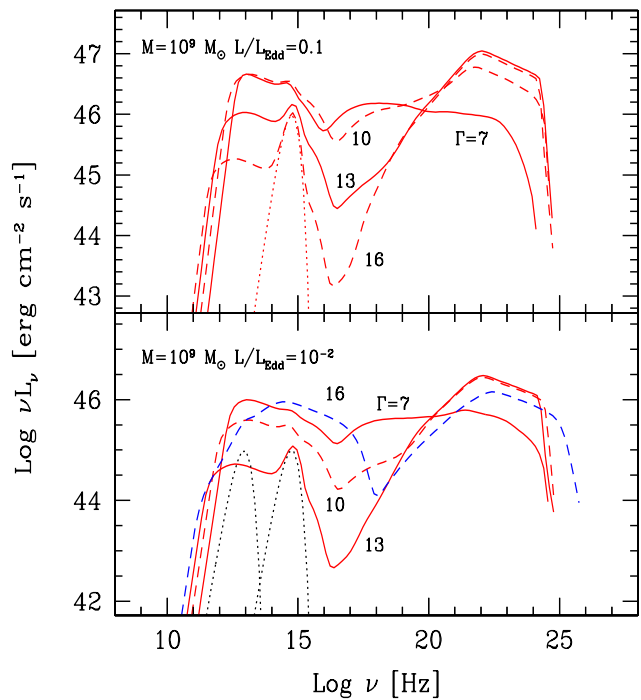,width=9.5cm,height=11cm} }
\vskip -0.8 true cm
\caption{
Predicted SED as a function of the bulk Lorentz factor
for a fixed black hole mass ($10^9 M_\odot$) and for 
$L/L_{\rm Edd}=0.1$ (top panel) and
$L/L_{\rm Edd}=10^{-2}$ (bottom panel).
For the latter case, note the SED of the $\Gamma=16$ case,
remarkably different from the others.
In this case $R_{\rm diss}>R_{\rm BLR}$ and no external BLR photons
contribute to the radiative cooling.
There is instead a contribution coming from the torus.
The reduced cooling implies a large $\gamma_{\rm peak}$,
making the SED similar to the one of a blue BL Lacs.
However, the broad lines do exists, and the objects would be classified
as ``blue"  quasars  (with a synchrotron peak frequency 
$\nu_{\rm peak}\sim 10^{15}$ Hz).
}
\label{sedgamma}
\end{figure}

\section{Changing parameters}

In the previous section we have shown the SED expected for different
values of the mass and accretion rate. Here we briefly discuss the
consequences of changing some other input parameters of the model.

\begin{itemize}

\item Changing the bulk Lorentz factor $\Gamma$.
In Fig. \ref{sedgamma} we show the predicted SED as a function of the 
bulk Lorentz factor for $L/L_{\rm Edd}=0.1$ (top panel) and
$L/L_{\rm Edd}=10^{-2}$ (bottom panel).
The black hole mass is kept constant ($M=10^9$ $M_\odot$).
The bulk Lorentz factors are in the range 7--16.
It can be seen that even small changes of $\Gamma$ imply
large changes in the observed SEDs, as described by Eq. \ref{lcls},
as long as $R_{\rm diss} < R_{\rm BLR}$.
Since we have assumed $a(\Gamma) \propto \Gamma^2$ 
for this test, we have $L_{\rm c}/L_{\rm s} \propto \Gamma^8$.
The top panel shows a monotonic sequence, since for all $\Gamma$
we have $R_{\rm diss} < R_{\rm BLR}$.
The bottom panel, instead, shows the case of a ``blue" quasar:
for the largest adopted $\Gamma$, in fact, we have $R_{\rm diss} > R_{\rm BLR}$,
and the external photons contributing to the EC spectrum are
coming from the torus only, contributing to the comoving radiation energy
density according to Eq. \ref{utorus}.
This implies a slower electron cooling, a larger $\gamma_{\rm peak}$ and
consequently larger peak frequencies.
Observers would however detect the broad emission lines
in this blazars, classifying it as a ``blue" FSRQ,
with a synchrotron peak frequency $\nu_{\rm peak}\sim 10^{15}$ Hz.

\item Changing $\epsilon_{\rm e}$ and $\epsilon_B$.
There are two different regimes: when the external 
seed photons dominate, $\epsilon_{\rm e}$ controls
the total produced luminosity, but not the Compton dominance,
which is instead proportional to $\epsilon_{\rm B}^{-1}$
(see Eq	\ref{lcls}).
When instead the high energy spectrum is dominated by the SSC flux,
the Compton dominance is proportional to 
$(\epsilon_{\rm e}/\epsilon_{\rm B})^{1/2}$ 
(as in Gamma Ray Burst afterglows).
 
\item Changing $R_{\rm diss}$.
Although we always assume   $R_{\rm diss}\propto M$,
we can change the constant of proportionality [$a(\Gamma)$].
As a rule, if we are inside the BLR, a smaller $R_{\rm diss}$
means a larger magnetic field, and thus a smaller Compton dominance
(see Eq. \ref{ub}; Eq. \ref{lcls}).
Also, increasing $a(\Gamma)$, makes the case $R_{\rm diss}>R_{\rm BLR}$
more probable, enlarging the parameter space for blue quasars.

\item Changing $s$. 
Besides changing the predicted slope above the peaks,
it changes $\epsilon_{\rm A}$ for objects emitting in slow cooling:
the steeper $s$, the less efficient the source (small $\epsilon_{\rm A}$).
Since the majority of blazars emitting in the slow cooling regime
are the ones with no BLR (and torus),
a steeper $s$ also means less (self--) Compton dominance.

\item Changing $\eta_{\rm j}$. 
This parameter act as a normalisation: larger $\eta_{\rm j}$
means that the jet is more powerful overall.
Keeping all other parameter constant, increasing $\eta_{\rm j}$
increases the value of the magnetic field, and therefore 
it decreases the Compton dominance when the external seed photons
dominate the radiation energy density.
When they do not, changing $\eta_{\rm j}$ has no effect on the
Compton dominance.

\end{itemize}

\section{Caveats}

Our proposed scenario is necessarily highly idealised 
and is aimed at describing the {\it averaged} behaviours
of different classes of blazars and also of a single
blazars that can have different states.
In this section we therefore try to summarise a few caveats
that should be kept in mind.

\begin{itemize}

\item Leptonic model. Probably the most basic assumptions of ours is
that the emission comes from the acceleration and the radiative
cooling of leptons, and we have neglected the possible presence of
relativistic hadrons (Mannheim 1993, Aharonian 2000, M\"ucke et al
2003). While the variability patterns observed in some (admittedly, a
few) blazars seem to favour the leptonic model, the hadronic scenario is
not ruled out yet.

\item Single zone emission region.  We have assumed that most of the
SED is produced in a single region, characterised by a single valued
and uniform magnetic field, with a single particle distribution, and
so on.  This is the most commonly adopted simplification, that
received support from the few cases in which we could see coordinated
variability in different energy bands, but consider that i) there
surely are several emitting regions, contributing differently to the
entire SED.  The region producing the radio flux, for instance, is
surely much larger than what we consider here, and could contribute
(by its SSC flux) to the X--ray radiation; ii) the jet could be
structured, composed by a fast spine surrounded by a slower
sheat--layer, as envisaged in Ghisellini, Tavecchio \& Chiaberge
(2005). Some support in this sense for TeV BL Lac objects comes from
the fact that these sources have ``slow" VLBI knots, often moving
sub--luminally (Piner \& Edwards 2004, Piner, Pant \& Edwards 2008 and
references therein), and from direct observations of some radio
edge--brightening (Giroletti et al. 2004, 2006).

\item A different $R_{\rm diss}$ for BL Lacs.
Associated to the previous point is the possibility that not
all blazars have the same $a(\Gamma)$ appearing in the
relation $R= a(\Gamma) R_{\rm s}$.
We have in fact assigned, to BL Lac objects,
a different $a(\Gamma)$ than in FSRQs, but the real value could
be different from what we have adopted.

\item Different bulk Lorentz factors $\Gamma$ for different classes of
objects.  There is the need, for TeV BL Lacs, of a very large
$\Gamma$, larger than for other blazars.  The reasons are the very
short variability timescales observed for the TeV flux (3 minutes in
PKS 2155--304, Aharonian et al. 2007; see the implications in
Begelman, Fabian \& Rees 2008 and models in Ghisellini \& Tavecchio
2008); and the necessity, in single zone SSC models, to account for
the large separation, in frequency, of the two peaks of the SED
(e.g. Konopelko et al. 2003; Tavecchio, Maraschi \& Ghisellini 1998).

\item The relation $\gamma_{\rm peak}(U)$.
This seems well obeyed by all observed sources of low--medium
power, while for powerful objects we have a larger scatter.
There is a branch at large $U$ for which $\gamma_{\rm peak}\propto U^{-1}$
instead of $\propto U^{-1/2}$ (see Celotti \& Ghisellini 2008).

\item Thomson vs Klein--Nishina scattering regime.
Our estimates assume that the inverse Compton process occurs in the 
Thomson regime, so only for $\gamma_{\rm peak}$ not extremely large.
We can estimate how large by the following argument.
If the EC process is important, then cooling is efficient, 
$\gamma_{\rm peak}< 10^3$--$10^4$ and the bulk of scatterings
indeed occurs in the Thomson regime.
When instead the SSC process is the main producer of high energy photons,
then we may neglect Klein Nishina effects when
$\gamma_{\rm peak}h\nu'_{\rm s} /(m_{\rm e} c^2) <1$, where
$\nu'_{\rm s}\sim 3.6\times 10^6 B\gamma^2_{\rm peak}$ 
is the synchrotron peak frequency in the comoving frame.
Therefore the Klein--Nishina regime is unimportant if
\begin{equation}
\gamma_{\rm peak}\, < \, \left( {m_{\rm e}c^2 \over 3.6\times 10^6 B h}
\right)^{1/3} \, \sim \, 3.2\times 10^4 B^{-1/3}
\end{equation}
This corresponds to observed synchrotron peak frequencies
$\nu_{\rm s}\sim 3.7\times 10^{15}B^{1/3}\delta$ Hz and self
Compton frequencies $\nu_{\rm c}= \gamma^2_{\rm peak}\nu_{\rm s}\sim
3.8\times 10^{24}B^{-1/3}\delta$ Hz $\sim 15.6B^{-1/3}\delta$ GeV.
With $\delta\sim 10$ and $B\sim 0.1$ G, blazars with Compton
peak energies below $\sim 70$ GeV should still be in the Thomson limit.

\item SED based on flares.
Our scenario is inevitably built upon what we already know, and
this is mostly limited by the sensitivity of EGRET.
Therefore, the complete SED we could construct, from the radio to the
GeV band, are most likely associated to sources in very active 
($\gamma$--ray) states, and might not be indicative of the more
standard, but more persistent, blazar spectrum.

\item Viewing angle. The models presented here assume that the jet is
observed with a small angle $\theta\sim 1/\Gamma\sim 4$ deg. According to the
current unification schemes, besides these highly beamed sources a
large number of moderately less beamed blazars should exist
($\theta\sim$4--7 deg), for which the non--thermal continuum of the jet is
less amplified and thus less luminous. An effective way to distinguish
between these sources and true low power FSRQs is through the level of
the optical emission, that in the misaligned objects should be
dominated by the ``naked'' thermal disk emission (see also Landt et
al. 2008).

\end{itemize}

\section{Implications for cosmic evolution}

The association of different classes of blazars with the mass
of the black hole and with the accretion rate
can have important effects on the predicted evolution of blazars.

Black hole of large masses ($M>10^9 M_\odot$) are rare,
while black holes of smaller masses are more common (at any redshift).
If the accretion rate in blazars evolves in redshift, as indicated
by the redshift evolution of radio--quiet quasars, we should have
larger accretion rates in the past.
As a consequence, blazars whose accretion rate is less than 
a critical value -- sources that we associate with BL Lac objects --
should become rarer in the past, implying a negative evolution.
On the contrary, FSRQs should be more common in the past, since
a larger fraction of massive black holes accretes at rates larger
than the critical one.

% The presence of the emission lines flags the existence of 
% a ``standard" disk, the absence should be associated to an ADAF--like
% accretion.

% Call $\dot m$ the accretion rate in units of the Eddington one, and
% put (arbitrarily) the division line between ADAF--like and Shakura
% Sunyaev accretion disk at $\dot m=10^{-2}$.

% BL Lacs with no broad emission lines are associated with
% $\dot m <10^{-2}$.
% FSRQs instead have $\dot m >10^{-2}$.

% Given a black hole mass, $\dot m$ is decreasing with cosmic time.

% Assume that the jet power (bulk kinetic+Poynting) is 
% always proportional to $\dot m$.
At large redshifts, when the black hole is fully
accreting, we have powerful disk and jets and
a BLR located at relatively large distances. 
These are the conditions to have a powerful, Compton dominated, FSRQ.
The same object will lower its accretion rate with time,
the disk becomes a less efficient radiator, the BLR
shrinks, and there should be the possibility to have
a transitional object, with emission lines of moderate 
luminosity, and with a SED that can  be produced sometimes 
within, sometimes outside, the BLR region (i.e. the red/blue FSRQ case
in Fig. \ref{plane}).
When the accretion rate decreases below the critical value
(i.e. $L_{\rm i}/L_{\rm Edd}<3\times 10^{-3}$, 
the accretion changes mode, becoming radiatively inefficient.
The corresponding BLR (if it exists at all) becomes weak and
very close to the black hole.
Jet dissipation occurs always beyond the BLR, and the source 
becomes a BL Lac object.
Synchrotron and SSC are the main emission processes, and the position
of the corresponding peaks in the SED depends on the amount of 
radiative cooling (bluer for less cooling).

These simple ideas are in line with the scenario proposed by
Cavaliere \& D'Elia  (2002), in which blazars are born initially
powerful (high accretion rate), and transform themselves into BL Lac
objects when the accretion rate decreases.
The evolution of blazars as a whole should be the
same of radio--quiet quasars (Maraschi \& Rovetti 1994), 
but with black hole masses
larger than $\sim 3\times 10^7 M_\odot$ (if this is the real
limit to have a radio loud object). 
SED--wise, the evolution is more complicated, since
for low masses and intermediate accretion rate, a FSRQ
can be blue ($R_{\rm diss}>R_{\rm BLR}$), but still have
visible broad lines.

\section{Conclusions}

We have proposed an extension of the scheme based on the so--called
``blazar sequence'', taking advantage from the new facts and ideas
discovered or suggested since the ``old" blazar sequence was
published.  Our aim is not to propose a new model or scenario for
blazars, but rather to view them from a different point of view,
enabling a different perspective.  Technically, we pass from a
one-parameter (observed bolometric luminosity) to a two-parameter
sequence ($M$ and $\dot M_{\rm in}$), but, more physically, we try to
link the properties of the jet to the main properties of the accretion
disk.  To do this, we had to make a number of simplifying assumptions,
that we think are rather reasonable or even supported by strong
observational evidence, but of course there remains some
uncertainties.  In addition to this, we stress that, at this stage, we
can only (and necessarily) describe {\it averaged properties} of
blazars and some {\it trends}.  But the link between the two
fundamental parameters ($M$ and $\dot M_{\rm in}$, or, rather, $L_{\rm
disk}$ and $M$) characterising the new sequence and the blazar SED
makes our proposal perfectly testable (and falsifiable).  The main
results are:

\begin{itemize}

\item
Blue quasars may exist.
If the dissipation region of the jet is beyond the broad line region, the
radiation energy density seen in the comoving frame is not large,
resulting in less electron cooling, and thus large electron energies.
This in turn results in large synchrotron and inverse Compton peak frequencies.
This might happen for large black hole masses and intermediate
accretion rates (for small $\dot M$ there are no broad lines anymore, and the 
objects becomes a BL Lac, for large $\dot M$ the radius of the BLR
is large, making it difficult that the jet produces most of its
luminosity beyond it).
If the dissipation region is related to the value of the bulk Lorentz
factor, as in the internal shock scenario, then an additional
requirement is that $\Gamma$ should be relatively large.
Note that the region of the parameter space corresponding
to blue quasars is not large, and its extension depends
upon $\Gamma$ and the relation between the radius of the BLR and
the disk luminosity. 
At this stage, it is then difficult to quantify more precisely
how many blue emission line blazars there are.
 
\item Low power blazars (red FSRQs; red and blue BL Lacs) should exist. 
They correspond to blazars associated with low mass black holes.
The blue BL Lacs, corresponding
to smaller accretion rates and hence smaller jet powers, can be missed
more easily (than red blazars) in flux limited samples.
It is possible that 
this effect can account for the large number of red blazar 
in deep flux limited samples (see e.g. Padovani et al. 2007)
especially if blazars with black holes of small masses are more numerous.
To demonstrate this, we of course need detailed simulations, that
we leave for future work.

\item Red low power blazars should exist.  Again they simply correspond
to objects with black holes of relatively small masses and relatively
large accretion rates, hence with a ``normal" broad line region.
% Again it is difficult to quantify how many red and weak blazars we
% expect, mainly because we do not know the low end of the black hole
% mass distribution of radio loud objects. 
Low power red sources should be
easily distinguished from the powerful misaligned counterparts
(expected to have a similar SED) since in these latter objects the
powerful disk emission is expected to dominate over the non-thermal
continuum in the optical-UV region (see Landt et al. 2008).

\item Not only the power of jets, but also the SED that
they produce is linked with $M$ and $\dot M_{\rm in}$.
This is the most important result of our study.

\end{itemize}

To confirm or falsify our ideas we need to know the SED, the mass of the black hole,
and the disk luminosity (or, equivalently, the luminosity of the broad 
emission lines).
We hope to soon enter in the ``{\it GLAST} era", greatly helping to 
determine the blazar bolometric luminosity and their high energy SED.
If the blazar sample of EGRET counts $\sim$60 objects, a improvement of a factor 10
in sensitivity means roughly that {\it GLAST} will detect $\sim 2000$ blazars.
Assuming that {\it Swift} can observe, simultaneously and with its optical and
X--ray telescopes, 1/10 of them, we can have a sample of $\sim$200
blazars with an optical to GeV {\it simultaneous} coverage.
This means a large improvement in number, and a huge improvement in quality,
of our knowledge.
These blazars can form a ``golden" sample for ground based follow up, aiming to
measure (if there are) the luminosity the broad emission lines and estimate
the mass of the central black hole.
With these information, 
we can easily confirm or reject the new blazar sequence we propose.

\section*{Acknowledgements}
We thank Annalisa Celotti and Laura Maraschi for useful discussions,
and the anonimous referee for helpful comments.

% ====================================================================
% ====================================================================

\newpage

\vskip 1 true cm
\noindent
{\large{\bf Appendix}}
\vskip 0.3 true cm

\noindent
{\large{\bf{The Comptonization $y$ parameter}}
\vskip 0.3 true cm

\noindent
Our method uses the fact that we ``know" the total radiation energy density
produced in the dissipation region, since it corresponds to the fraction
of energy lost by the electrons in one light crossing time:
$U^\prime_{\rm r}=\epsilon_{\rm A} U_{\rm e}$.
This radiation energy density $U^\prime_{\rm r}$ 
is the sum of the synchrotron and inverse Compton 
emission\footnote{$U^\prime_{\rm r}$ should not be confused with
$U^\prime$ of Eq. \ref{gpeak}, which is the magnetic plus 
radiative energy density available for efficient scattering and
contributing to the cooling.}.

Consider first the synchrotron emission, with a corresponding 
cooling rate given by 
$ \dot \gamma_{\rm s} = (4/3)\sigma_{\rm T} c U_B \gamma^2/(m_{\rm e} c^2)$.
Neglecting self--absorption, the total (comoving) synchrotron luminosity is
\begin{equation}
L^\prime_{\rm s}\, 
=\, \pi r^2\Delta R^\prime m_{\rm e} c^2 \int N(\gamma) \dot\gamma_{\rm s} d\gamma
\label{lsyn}
\end{equation}
We now divide and multiply  by $n \equiv \int N(\gamma)d\gamma$ and define
$\tau\equiv n \Delta R^\prime n$.
We can also write $L^\prime_{\rm s}$ as $U^\prime_{\rm s}  \pi r^2 c$.
We then have
\begin{equation}
U^\prime_{\rm s} \, =\, U_B \, {4\over 3} \tau \langle \gamma^2 \rangle \, \equiv \, y U_B
\label{lsyn2}
\end{equation}
where the last equivalence corresponds to define the Comptonization parameter 
$y\equiv (4/3) \tau \langle \gamma^2 \rangle $.
Assuming that the scattering process occurs in the Thomson regime
we can analougously set
$U^\prime_{\rm SSC} = y U^\prime_{\rm s} =y^2 U_B$, and 
$U^\prime_{\rm EC} = y U^\prime_{\rm ext}$.
The sum of these components constitutes $U^\prime_{\rm r}$:
\begin{equation}
U^\prime_{\rm r}  = U^\prime_{\rm syn} +U^\prime_{\rm SSC} + U^\prime_{\rm EC} =
yU_{\rm B} + y^2 U_{\rm B} + y U^\prime_{\rm ext} 
\label{y}
\end{equation}
Dividing both parts by $U_{\rm B}$ we obtain an equation 
dependent on the microphysical parameters $\epsilon_B$, $\epsilon_{\rm e}$
and $\epsilon_{\rm A}$ (defined in \S 2.7, see Eqs. \ref{ub}, \ref{ue}), and dependent 
on $L_{\rm j}$ only through the ratio $U^\prime_{\rm ext}/ U_B$:
\begin{equation}
{U^\prime_{\rm r} \over U_B} \, =\, {\epsilon_{\rm A} \epsilon_{\rm e} \over \epsilon_B} 
\, =\,  y^2  + y \left( 1+  {U^\prime_{\rm ext} \over U_B} \right)
\end{equation}
The solution is:
\begin{equation}
y\, =\, {1\over 2} \, \left\{ \left[ 
\left( 1+ { U^\prime_{\rm ext}\over U_{\rm B} }\right)^2 
+{ 4\epsilon_{\rm e}\epsilon_{\rm A} \over \epsilon_{\rm B} } \right]^{1/2} -
\left( 1+ { U^\prime_{\rm ext}\over U_{\rm B} }\right) \right\}
\label{ysolved}
\end{equation}
For $U^\prime_{\rm ext}/U_{\rm B}$ much smaller or greater than unity
we have the limiting cases of Eq. \ref{y_array}.

\vskip 0.5 true cm
\noindent
{\large{\bf Finding $\epsilon_{\rm A}$}}
\vskip 0.5 true cm

\noindent
To find $\epsilon_{\rm A}$, consider
an injected relativistic electron distribution that has a power law form, 
$Q(\gamma) =Q_0 \gamma^{-s}$, between some $\gamma_1$ and $\gamma_2$.
The fraction $\epsilon_{\rm A}$ of radiated energy (in one light crossing time) 
can be approximated as
\begin{equation}
\epsilon_{\rm A} \, =\, 
{\int_{\gamma_{\rm peak}}^{\gamma_2} \gamma^{1-s} d\gamma \over
\int_{\gamma_1}^{\gamma_2} \gamma^{1-s} d\gamma } \, \sim \, 
\left( \gamma_1 \over \gamma_{\rm peak}\right)^{s-2}; \,\,\, s>2
\label{epsa} 
\end{equation}
The value of $\gamma_{\rm peak}$ depends upon 
$U^\prime = U_B +^\prime_{\rm s} +U^\prime_{\rm ext}$ (Eq. \ref{gpeak}).
In particular, the synchrotron radiation energy density depends upon $y$
(Eq. \ref{y}), i.e. it depends on $\epsilon_{\rm A}$ itself.
Despite this circularity problem, we can solve the system iteratively,
until convergence (obtained in a few cycles).

\end{document}